%% file: main.tex
\documentclass[reprint,aps,pre]{revtex4-1}

\usepackage{amsmath,amsfonts,amscd,amssymb,graphicx,subeqnar}

\input{z_vardef}
\usepackage{dcolumn}
\usepackage{overpic}
\usepackage{color}
\usepackage{adjustbox}
\usepackage[caption=false]{subfig}
\usepackage{rotating}
\usepackage{float}
\makeatletter
\let\newfloat\newfloat@ltx
\makeatother
\usepackage{algorithm}
\usepackage{algpseudocode}
\algnewcommand\algorithmicinput{\textbf{Input:}}
\algnewcommand\Input{\item[\algorithmicinput]}

\begin{document}

\title{Shallow Recurrent Decoder for Reduced Order Modeling of Plasma Dynamics}

\author{J. Nathan Kutz$^{\dag}$, Maryam Reza$^{*}$,  Farbod Faraji$^{*}$, Aaron Knoll$^{*}$}
 \affiliation{$^\dag$Department of Applied Mathematics and  Electrical and Computer Engineering, University of Washington, Seattle, WA 98195}
\affiliation{$^{*}$Imperial Plasma Propulsion Laboratory, Department of Aeronautics, Imperial College London, London, UK}

\begin{abstract}
Reduced order models are becoming increasingly important for rendering complex and multiscale spatio-temporal dynamics computationally tractable.  The computational efficiency of such surrogate models is especially important for design, exhaustive exploration and physical understanding.  Plasma simulations, in particular those applied to the study of ${\bf E}\times {\bf B}$ plasma discharges and technologies, such as Hall thrusters, require substantial computational resources in order to resolve the multidimentional dynamics that span across wide spatial and temporal scales.  Although high-fidelity computational tools are available to simulate such systems over limited conditions and in highly simplified geometries, simulations of full-size systems and/or extensive parametric studies over many geometric configurations and under different physical conditions are computationally intractable with conventional numerical tools.  Thus, scientific studies and industrially oriented modeling of plasma systems, including the important ${\bf E}\times {\bf B}$ technologies, stand to significantly benefit from reduced order modeling algorithms. We develop a model reduction scheme based upon a {\em Shallow REcurrent Decoder} (SHRED) architecture. The scheme uses a neural network for encoding limited sensor measurements in time (sequence-to-sequence encoding) to full state-space reconstructions via a decoder network.  Based upon the theory of separation of variables, the SHRED architecture is capable of (i) reconstructing full spatio-temporal fields with as little as three point sensors, even the fields that are not measured with sensor feeds but that are in dynamic coupling with the measured field, and (ii) forecasting the future state of the system using neural network roll-outs from the trained time encoding model.  The SHRED reduced order model architecture is demonstrated on the plasma dynamics in a 2D configuration representative of a radial-azimuthal plane of a typical Hall thruster geometry.

\end{abstract}

\maketitle

\section{Introduction}

Plasma physics simulations are of critical importance for characterizing the underlying multiscale plasma dynamics for a given geometry and experimental configuration. Indeed, due to the complexity of measuring quantities of interest in laboratory settings, computational studies can be highly beneficial to assess the viability of plasma technologies as well as to aid the fundamental understanding of the underlying physics of operation. Nonetheless, resolving the multiscale physics and the intricate interplay of various phenomena across multiple dimensions makes traditional high-fidelity plasma simulations computationally intensive and/or prohibitive due to their high dimensionality, especially when considering full three-dimensional geometries of plasma systems, such as Hall thrusters ~\cite{Powis_2023,Villafana_2023,Reza_2023}, Penning discharges ~\cite{Powis_2023}, and fusion configurations, for instance, the z-pinch~\cite{ZPinch_2001}. Reduced order models (ROMs)~\cite{benner:2015,brunton2019data,kutz:2013} have emerged as a viable mathematical architecture for producing surrogate models that can significantly accelerate computational studies of high-dimensional and coupled spatio-temporal systems.  We introduce the {\em Shallow REcurrent Decoder} (SHRED)~\cite{williams2022data,ebers2023leveraging} architecture as a deep learning {\em reduced order model} (ROM) for ${\bf E} \times {\bf B}$ plasma dynamics, showing that the learned SHRED model provides an accurate proxy for the high-fidelity simulations at a comparatively negligible computational cost.  Moreover, the SHRED model can be trained on compressive representations of the coupled spatio-temporal fields that characterize the plasma dynamics, allowing for the measurement of a single quantity of interest and the  reconstruction of all other spatio-temporal plasma fields.

\begin{figure*}[t]
   \centering
  \begin{overpic}[width=0.8\textwidth]{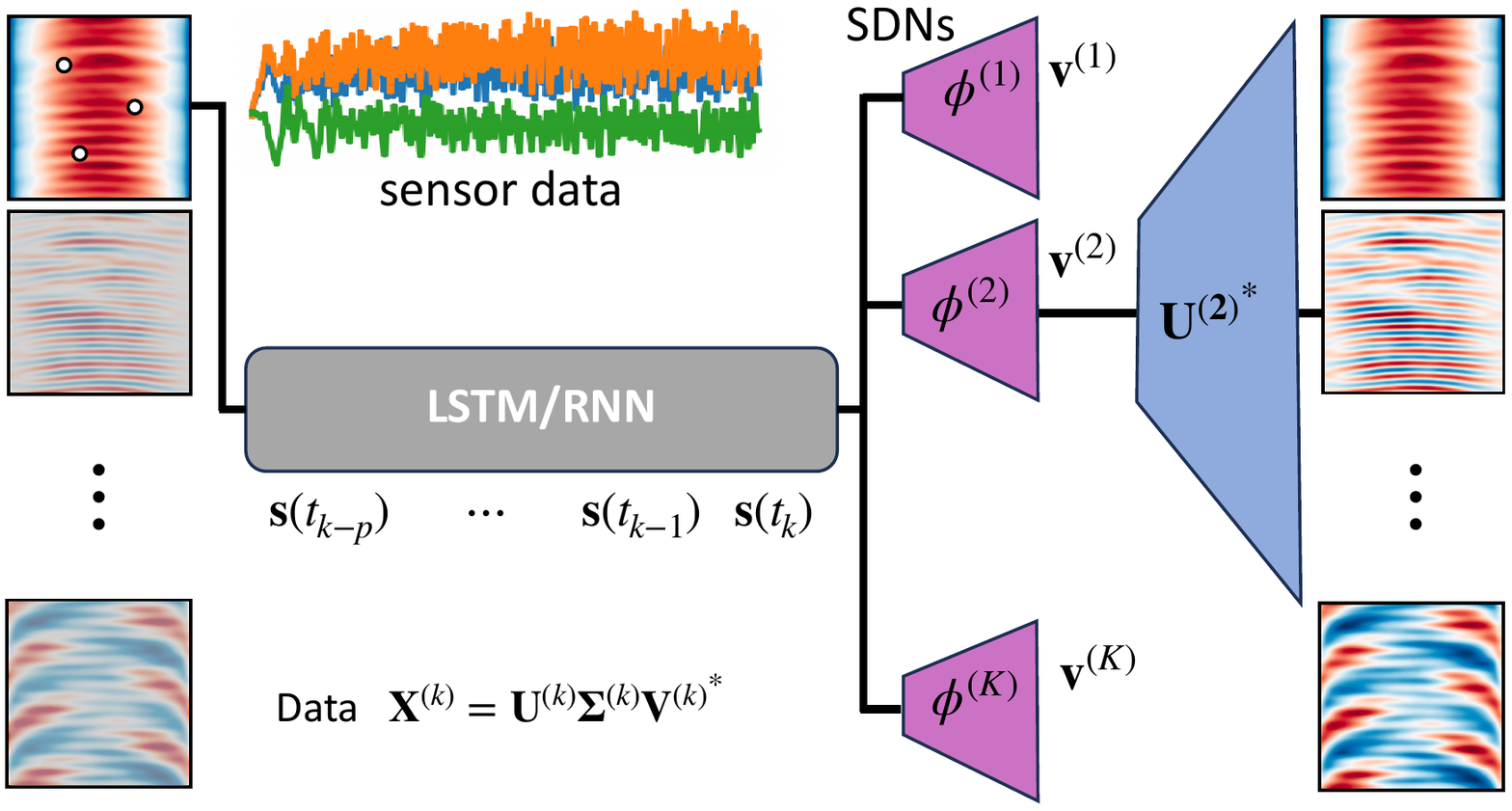}
\put(-2,58){${\bf n}_e$}
\put(100,58){${\bf n}_e$}
\put(100,45){${\bf E}_z$}
\put(100,19){${\bf J}_{ey}$}
  \end{overpic}
  \vspace*{-.4in}
   \caption{Architecture of the SHRED model for emulating plasma dynamics.  A single field is measured, specifically ${\bf n}_e$, using sensor history (trajectory) data and the model is trained to map to the original data of the fourteen fields.  In this case, the SHRED model is trained specifically to map to the compressive representation of the full plasma dynamics by mapping to the $r$-rank right singular values (${\bf V}^{(k)}$) of a given field computed by a randomized singular value decomposition ${\bf X}^{(k)} = {\bf U}^{(k)} {\bf \Sigma}^{(k)} {{\bf V}^{(k)}}^*$. Reconstruction of the $k$th field can be accomplished by projecting to the high-dimensional space using ${{\bf U}^{(k)}}^*$. }
   \label{fig:Figure1}
\end{figure*}

The SHRED architecture is an example of a data-driven model that is constructed through a training and learning process.  It is a generalization of the separation of variables method for solving partial differential equations (PDEs)~\cite{williams2022data,ebers2023leveraging}.  Data-driven models have recently emerged as a leading paradigm for characterizing complex dynamics.  Models such as {\em dynamic mode decomposition} (DMD)~\cite{kutz2016book,ichinaga2024pydmd} and its variants, especially the optimized~\cite{askham2018variable} and bagging optimized DMD (BOP-DMD)~\cite{sashidhar2022bagging} are simple, robust and stable regression techniques that have proven effective in modeling plasma dynamics~\cite{taylor2018dynamic,kaptanoglu2020pop,faraji2023dynamic,faraji2023dynamic2}.  The {\em sparse identification of nonlinear dynamics} (SINDy)~\cite{brunton2016extracting,kaptanoglu2021pysindy,nicolaou2023data} similarly allows for learning nonlinear proxy dynamics of plasmas~\cite{kaptanoglu2020physics,lore2023time} through a sparse regression procedure that can be made robust~\cite{fasel2021ensemble,gao2022bayesian,messenger2021weak}.  The novel data-driven local-operator finding algorithm, Phi Method, enables a robust, simultaneous discovery of the nonlinear dynamics and the optimal descretization stencil for the involved dynamics variables via constrained regression \cite{farajiMLSciTech,faraji2024PhiMethodI,faraji2024PhiMethodII}. The advantage of all these methods is that interpretable proxy models can approximate the plasma dynamics at a fraction of the computational cost of full simulations.

SHRED is a deep learning model that exploits the separation of time and space to build a reduced order, nonlinear proxy model.  Figure~\ref{fig:Figure1} highlights the architecture that will be advocated in what follows. SHRED combines a recurrent neural network for modeling the time dynamics of a small number of sensors with a decoder network that models the full spatial map of the involved quantities. Combined, they provide a nonlinear generalization of separation of variables~\cite{williams2022data,ebers2023leveraging}.  The recurrent neural network~\cite{lipton2015critical}, which is chosen to be an LSTM (long short-term memory)~\cite{hochreiter1997long,yu2019review}, models a time sequence of measurements, or trajectory, from a limited number of point sensors in any one of the spatio-temporal fields of the plasma dynamics.  The LSTM itself constructs a latent space representation of the dynamics given a time-lagged embedding.  Such time-lagged embeddings have been shown to be related to Takens embedding theory~\cite{takens1981lnm}.  Moreover, it has been exploited in both DMD~\cite{brunton2017natcomm,arbabi2018cdc} and SINDy~\cite{bakarji2023discovering,rosafalco2024ekf,leylaz2022identification}.  The latent space then projects the data through a shallow decoder network (SDN)~\cite{erichson2020shallow} back to the high-dimensional state space of the coupled plasma fields.  As shown in Fig.~\ref{fig:Figure1}, we can even train the shallow decoder to map to a compressed space, for example, the low-rank space spanned by the SVD. This compressed training is then used to map back to the full state space through the $r$-rank left singular eigenvectors of the SVD, ${\bf U}^*$.  Thus the model can be trained on laptop level computing platforms.

As will be shown, SHRED has a theoretical basis in the PDE  theory of separation of variables.  It exploits the fact that two coupled first order PDEs can be written as a second order PDE in a single variable.  Or more generically, $N$ coupled PDEs can be written as an $N$th order PDE in a single variables.  Specifically, SHRED can be rigorously justified for linear PDEs, and the SHRED architecture is then overall a generalization to nonlinear PDEs.  The advantages of the SHRED architecture are many, including (i) the ability to use only three sensors for reconstructing the entire plasma dynamics, (ii) the ability to train on compressed data, (iii) the ability to measure a single field and reconstruct coupled spatio-temporal fields that are not measured (sensed), and (iv) minimal hyper-parameter tuning. 

The SHRED architecture is agnostic toward sensor placement, which is specifically important for plasma systems since most of the plasma field might be inaccessible to the probes/diagnostics. Furthermore, SHRED is uniquely advantageous for plasma systems' analysis and modeling also because it can enable full-state reconstruction/forecasting with measurements of a single quantity that might be the simplest (least expensive) to diagnose.

We demonstrate the SHRED model as a ROM on simulations of complex, highly coupled plasma dynamics in a Hall thruster-representative configuration, showing its reconstruction properties and forecasting properties.  The observed performance makes SHRED a valuable tool for emulating plasma physics at a negligible computational cost.

\section{Overview of ${\bf E} \times {\bf B}$ Plasmas}

"Cross-field" ${\bf E} \times {\bf B}$ plasmas consist of an often partially magnetized plasma immersed in a perpendicular orientation of the electric (${\bf E}$) and the magnetic (${\bf B}$) fields. The term "cross-field" stems from this mutually perpendicular fields' configuration.

${\bf E} \times {\bf B}$ plasma technologies have found industrially crucial and important applications. These applications include the magnetrons, which are devices that are central in the manufacturing industry, especially toward development of (silicon) microchips for computer processors. Another prominent example of the industrial applications of ${\bf E} \times {\bf B}$ plasmas is the Hall thrusters for spacecraft propulsion. Hall thrusters are today the most widely in-use electric propulsion solution for satellites. They are also prime candidates to enable the next generation of near-Earth and interplanetary space missions. 

Besides their applied significance, ${\bf E} \times {\bf B}$ plasmas have received remarkable scientific attention over the past decades \cite{ExBPlasmas}. This is partly because cross-field plasma discharges exhibit a complex and rich underlying physics that, in certain respects, resemble the phenomena occurring in the other areas of applied plasma physics, such as fusion energy. Nevertheless, the ${\bf E} \times {\bf B}$ plasma technologies are more accessible to study and much less costly to test than the fusion reactors. Therefore, investigating the plasma phenomena in ${\bf E} \times {\bf B}$ devices can serve as a cost-effective approach for deriving useful research insights into fusion dynamics. 

In addition, the operation of ${\bf E} \times {\bf B}$ technologies is strongly influenced by the underlying plasma processes. Most notably, the efficiency of an ${\bf E} \times {\bf B}$ device is largely dependent on the effectiveness with which the applied magnetic field can hinder the current of the electron species across the magnetic field lines. However, there are a myriad of physical phenomena that “undesirably” [from an applied perspective] enhance the cross-magnetic-field motion (transport) of the electrons, most of which are yet not fully understood \cite{ExBPlasmas, ExBPhysics}. This close coupling between the engineering and the physics of operation of ${\bf E} \times {\bf B}$ technologies gives the scientific research into these devices a critical applied importance. This has been another main driver of academic interest in investigating plasmas in ${\bf E} \times {\bf B}$ devices.

In order to study many of the yet-unresolved plasma physics questions, and also for the endeavors aimed at realizing predictive plasma models \cite{lafleur2016, mikellides2016, reza2017, jorns2018, marks2023}, the Hall thruster configuration has presented itself as a suitable proving ground over the past years.  Indeed, the laboratory-scale ${\bf E} \times {\bf B}$ plasma in a Hall thruster exhibits several important processes and behaviors that are present in other plasma regimes as well, such as in fusion plasmas and in space plasmas. Moreover, a Hall thruster offers a relatively convenient accessibility and affordability for numerical and experimental physics studies over other industrially relevant types of plasma sources while closely matching their plasma configuration. 

As shared among the broader family of ${\bf E} \times {\bf B}$ plasma discharges and technologies, Hall thrusters have a highly complex, multiscale, and multidimensional underlying physics \cite{boeuf2017, ExBPhysics}. The cross-field configuration of the plasma in Hall thrusters leads to strong anisotropies in the properties of the magnetized electrons with respect to the directions along and perpendicular to the magnetic field lines \cite{taccogna2019}. These anisotropies and the strong spatial gradients in the plasma properties result in the excitation of a diverse range of plasma instabilities and turbulence that span a broad range of spatial and temporal scales  \cite{choueiri2001, ExBPlasmas, boeuf2017}. As the instabilities excite, grow, and saturate, they can strongly interact with the plasma species and influence their spatio-temporal dynamics. The instabilities can also interact with each other, exchanging energy, and cause the formation of larger, possibly turbulent, global plasma structures \cite{koshkarov2019}.

According to this overview, the complexity of the physics in a Hall thruster, with phenomena and behaviors extending to across the broader field of plasma physics, implies that a reduced-order model (or modelling approach/methodology) capable of predicting the spatio-temporal evolutions of the plasma in this device in an accurate, robust, and generalizable manner effectively paves the way to achieve the long-sought-after predictive models across the plasma science domains. As a result, the demonstrations of SHRED in this work focuses on reconstruction and forecasting of plasma dynamics in a Hall-thruster-representative test problem. 

\section{Simulation setup and the computational tool for data generation}

The setup of the high-fidelity simulation used to generate data for SHRED's training and testing in this work resembles a 2D radial-azimuthal plane of a typical Hall thruster geometry. The simulation setup and conditions are identical to those of the radial-azimuthal benchmark problem defined by Villafana et al. and reported in Ref. \cite{villafana2021}.  This benchmark problem is a well-studied one. In particular, extensions to this problem were adopted in Refs. \cite{reza2023parametric, reza2023Influence, reza2024EffectsI, reza2024EffectsII} for in-depth and broad-scale parametric studies on the dominant and influential plasma phenomena and instabilities. 

The simulation plane is Cartesian, and the coordinate system comprises the $x$ and the $z$ axes. The $x$ coordinate is along the radial direction and the $z$ coordinate along the azimuthal direction. The simulation domain also features a fictitious axial ($y$ ) extent \cite{villafana2021}, which serves to limit the growth of the energy of the system \cite{lafleur2016} by mimicking the axial convection of the plasma species that occurs in a realistic Hall thruster geometry.  Further details of the simulation setup, including the physical and numerical parameters, are described at length in Refs. \cite{farajiVerification}.

High-fidelity particle-in-cell (PIC) simulations are performed for the adopted radial-azimuthal test problem using the 2D PIC code of Imperial Plasma Propulsion Laboratory, IPPL-2D \cite{farajiVerification}. IPPL-2D follows the standard implementations of electrostatic, explicit, momentum conserving PIC codes, is written in the Julia language \cite{bezanson2017}, and is parallelized with the Message Passing Interface (MPI) protocol. IPPL-2D was verified against the results of the radial-azimuthal benchmark \cite{villafana2021} in Ref. \cite{farajiVerification}.

The plasma in our test case here exhibits a complex behavior that is primarily influenced by two important instability modes, the electron cyclotron drift instability (ECDI) \cite{janhunen2018, ducrocq2006, cavalier2013, tsikata2010}, and the modified two-stream instability (MTSI) \cite{petronioMTSI, janhunenMTSI, reza2023parametric, Villafana_2023}. In this regard, following the nonlinear development of these instabilities during the transient phase of the discharge evolution, a quasi-steady state is reached characterized by the nonlinear interactions between the ECDI and the MTSI. This interaction leads to periodic mitigation and growth of these instabilities. The dynamic interplay between the ECDI and the MTSI is reflected in the time evolution of the macroscopic plasma properties, particularly the ion number density and the radial electron temperature, in terms of a cyclic rise and drop \cite{villafana2021, farajiMLSciTech}.

The complex nature of the problem and the dominant role of underlying instability modes in determining the global system response make the radial-azimuthal test case an interesting problem to assess the predictive performance of ROMs from the SHRED architecture. 

\section{SHALLOW RECURRENT DECODERS:  SHRED}

The SHRED architecture is based upon the separation of variables technique for solving linear partial differential equations (PDEs)~\cite{folland2020introduction}.  Separation of variables assumes that a solution can be separated into a product of spatial and temporal functions $u(x,t)= T(t) X(x)$.  This solution form is then used to reduce the PDE into a ordineary differetial equations:  one for time $T(t)$ and one for space $X(x)$.  Such a decomposition also constitutes the underpinnings of spectral methods for the numerical solution of the PDE, linear or nonlinear~\cite{kutz:2013}.  

Consider the constant coefficient linear PDE 
\begin{equation}
 \dot{u} = {\cal L} (\partial_x, \partial^2_x, \cdots ) {u}
 {\label{eq:linearPDE}}
\end{equation}
where $u(x,t)$ specifies the spatio-temporal field of interest. Typically the initial condition (IC) and boundary conditions (BCs) are given by
\begin{subequations}
\begin{align}
\text{IC:}&&\quad  u(x,0)=u_0(x) \\
\text{BCs:}&&\quad  \alpha_1 u (0,t) +\beta_1 u_x(0,t) = g_1(t) \,\, \text{and} \nonumber \\  && \alpha_2 u(L,t) + \beta_2 u_x(L,t) = g_2(t).
\end{align}
\label{eq:ICBC}
\end{subequations}
This may be generalized to systems of several spatial variables, or a system with no time dependence.  The linear operator ${\cal L}$ specifies the spatial derivatives, which in turn model the underlying physics of the system.  Simple examples of ${\cal L}$ include ${\cal L}= c \partial_x$ (the one-way wave equation) and ${\cal L}= \kappa \partial^2_x$ (the heat equation)~\cite{kutz:2013}.  

The earliest solutions of linear PDEs assumed separation of variables whereby $u(x,t)=\exp(\lambda t) X(x)$ was a product of a temporal (exponential) function multiplied by a spatial function.  The parameter $\lambda$ is in general complex.  This gives the eigenfunction solution of (\ref{eq:linearPDE}) to be
\begin{equation}
    u(x,t) = \sum_{n=1}^{N} a_n \exp(\lambda_n t) \phi_n(x)
    \label{eq:ef}
\end{equation}
where $\phi_n(x)$ are the eigenfunctions of the linear operator and $\lambda_n$ are its eigenvalues (${\cal L}\phi_n(x) = \lambda_n \phi_n(x)$).  Here a finite dimensional approximation $N$ is assumed, which is standard in practice for numerical evaluation.  

Typically, initial conditions $u(x,0)=u_0(x)$ are imposed in order to uniquely determine the coefficients $a_n$.  Specifically, at time $t=0$ (\ref{eq:ef}) becomes
\begin{equation}
    u_0(x) = \sum_{n=1}^{N} a_n  \phi_n(x) .
\end{equation}
Taking the inner product of both sides with respect to $\phi_m(x)$ and making use of orthogonality gives
\begin{equation}
   a_n =\langle u_0(x), \phi_n(x) \rangle
\end{equation}
Instead of having the full spatial distribution at $t=0$ ($u_0$), SHRED has the measurements at a single spatial (sensor) location $x_s$, but with a temporal history.  Thus if SHRED has, for example, $N$ temporal trajectory points, this gives at each time point of the measurement:
\begin{equation}
    u(x_s, t_j) = \sum_{n=1}^{N} a_n 
    \exp(\lambda_n t_j) \phi_n(x_s) 
    \,\,\,\,\,\, \mbox{for} \,\,\, j=1,2,\cdots N .
\end{equation}
This results in $N$ equations for the $N$ unknowns $a_n$. Specifically, the $N\times N$ system of equations ${\bf A} {\bf x} = {\bf b}$ is prescribed by the vector components $x_k=a_k$ and $b_k = u(x_s,t_k)$ and matrix components $(a_{kj}) = \exp(\lambda_k t_j) \phi_k(x_s)$.  As with the initial condition (\ref{eq:ICBC}a), the time trajectory of measurements at a single location uniquely prescribes the solution.  This analysis can easily be generalized to include multiple sensor measurements at a single time point. Thus if there are two measurements at a given time $t_j$, then only $N/2$ trajectory points are needed to uniquely determine the solution.  Likewise, three sensor measurements at a given time requires on $N/3$ trajectory points.
In addition to stationary sensors measurements, one can also consider mobile sensors whereby the measurement of the system is a different locations over time:  $x_s = x_{s(t_j)}$. The above arguments are easily modified so that the vector component $b_k = u(x_{s(t_j)},t_k)$ and matrix components $(a_{kj}) = \exp(\lambda_k t_j) \phi_k(x_{s(t_j)})$.

Thus, temporal trajectory information at a single spatial location, or with a moving sensor, is equivalent to knowing the full spatial field at a given point in time. SHRED is a generalization of the separation of variables architecture $u(x,t) = T(t)X(x)$ where the LSTM models time $T(t)$ and the decoder models space $X(x)$.  Of course, as a generalization to nonlinear dynamics, rigorous theoretical bounds of SHRED are difficult to achieve, much like analytic and numerical solutions are difficult to rigorously bound in computational PDE settings.  But certainly in the linear limit, the above arguments show explicitly why SHRED is guaranteed to work and recover the full spatio-temporal field exactly.

\subsection{Nonlinear PDEs}

The above arguments show that time and space measurements can be traded for each other explicitly for linear PDEs.  For nonlinear PDEs
\begin{equation}
 \dot{u} = N (u, u_x, u_{xx}, \cdots ), 
 {\label{eq:nonlinearPDE}}
\end{equation}
numerical solution techniques are typically used to generate solutions subject to the initial and boundary conditions~(\ref{eq:ICBC}).  Consider a spectral solution technique~\cite{kutz:2013} whereby numerical solutions are approximated by a spectral basis
\begin{equation}
    u(x,t) = \sum_{n=1}^{N} a_n (t) \phi_n(x) .
    \label{eq:spectral}
\end{equation}
Typical examples of spectral techniques include using Fourier modes or Chebychev polynomial for $\phi_n(x)$.  This spectral decomposition turns the PDE into a systems of $N$ coupled ordinary differential equations for $a_n(t)$:
\begin{equation}
  \frac{da_n}{dt}=f_n (a_1, a_2, \cdots, a_N) \,\,\,\,\,\, \mbox{for} \,\,\, n=1, 2, \cdots N 
  \label{eq:an}
\end{equation}
The solution of the $N$-dimensional differential equation has $N$ unknown constants of integration that are typically uniquely determined by applying initial conditions and orthogonality in (\ref{eq:spectral})
\begin{equation}
    a_n(0)= \langle u_0(x), \phi_n(x) \rangle   . 
\end{equation}
As with the separation of variables solution, we can instead assume that we can construct a general solution  for
(\ref{eq:an}) which has $N$ constants of integration. The constants of integration can be determined by requiring the solution to satisfy $N$ temporal trajectory points, giving at each time point of the measurement:
\begin{equation}
    u(x_s,t_j)=\sum_{n=1}^N a_n(t_j)\phi_n(x_s)  \,\,\,\,\,\, \mbox{for} \,\,\, j=1,2,\cdots N .
\end{equation}
This gives $N$ constraints for the $N$ unknown constants of integration, thus uniquely determining the evolution of the $a_n$ in (\ref{eq:an}).  Mobile sensors can also be used to enforce the constraints required for a unique solution.\\

\subsection{Coupled PDEs}

We can also consider coupled, constant coefficient linear PDEs.  For example, the coupled system 
\begin{subeqnarray}
 \dot{u} = {\cal L}_1  {u}
 + {\cal L}_2  {v}\\
 \dot{v} = {\cal L}_3  {u}
 + {\cal L}_4 {v}
 {\label{eq:linearPDE2}}
\end{subeqnarray}
where $u(x,t)$ and $v(x,t)$ specifies the spatio-temporal fields of interest. The PDEs can be instead be written in the form
\begin{equation}
    \ddot{u} = {\cal L}_1 \dot{u}
    +{\cal L}_2 {\cal L}_3 u
    +{\cal L}_2 {\cal L}_4 \left(
     {\cal L}_2^{-1} (\dot{u} - {\cal L}_1 {u})
    \right)
\end{equation}
where (\ref{eq:linearPDE2}a) is differentiated with respect to time and (\ref{eq:linearPDE2}b) is used in order to write the PDEs as a function of $u(x,t)$ alone.  Thus knowledge of the field $u(x,t)$ alone is capable of constructing the solution fields $u(x,t)$ and $v(x,t)$.  For this second-order (in time) PDE, both an initial condition $u(x,0)$ and an initial {\em velocity} require specification 
$\dot{u}(x,0)$ in order to uniquely determine the solution.  As with the previous arguments, a time trajectory embedding of $2N$ measurements can be used to uniquely determine the solution.
 
\begin{figure}[t]
   \begin{overpic}[width=0.45\textwidth]{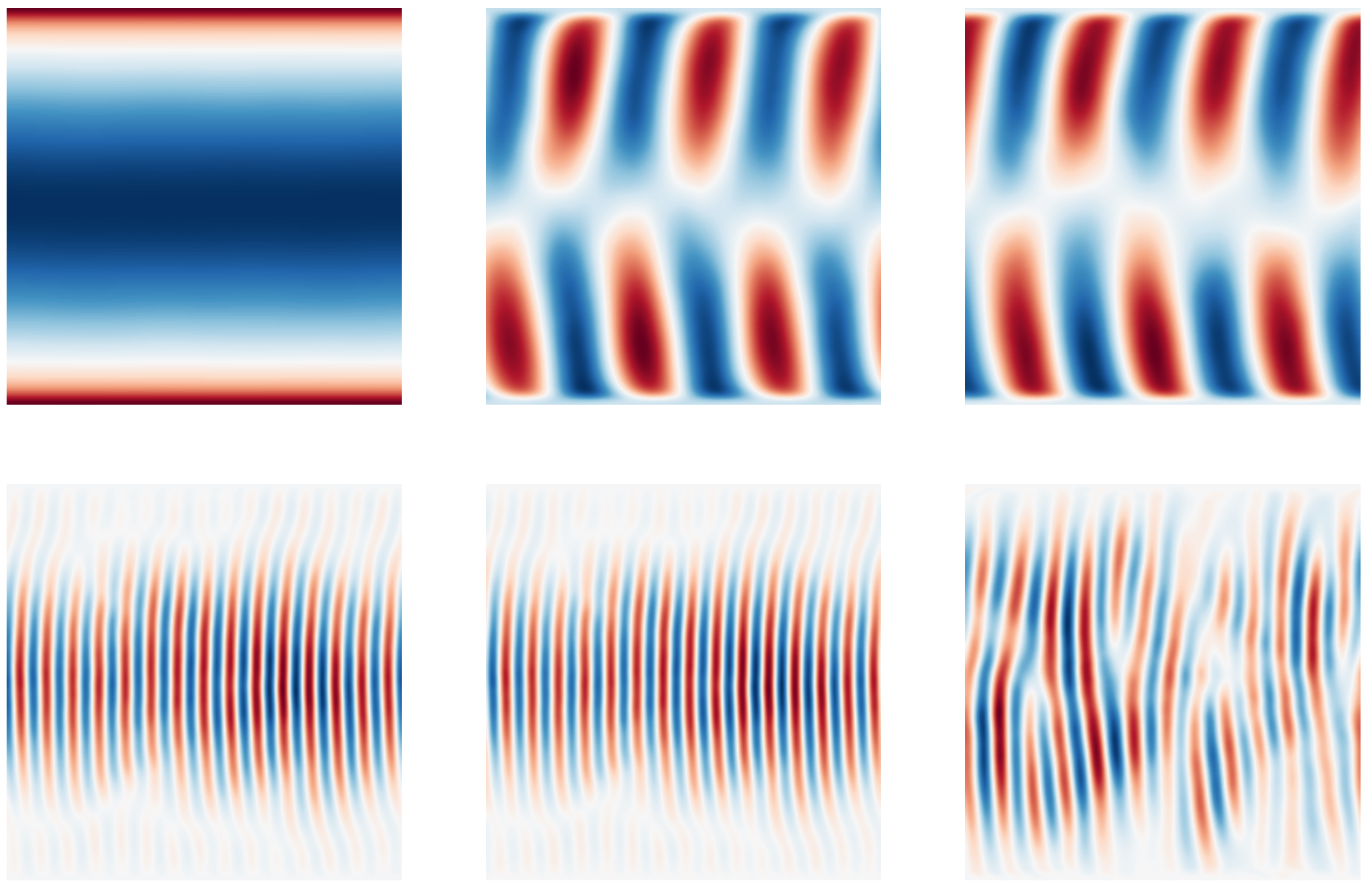}
\put(1,66){${\bf u}_1$}
\put(36,66){${\bf u}_2$}
\put(70,66){${\bf u}_3$}
\put(1,31){${\bf u}_4$}
\put(36,31){${\bf u}_5$}
\put(70,31){${\bf u}_6$}
   \end{overpic}
 \caption{First six principal components (SVD modes ${\bf u}_k$ for $k=1, 2, 3, 4, 5, 6$) of the ${\bf n}_e$ dynamics.  The resolution of the simulation is $n_x=256 \times n_z=257$ in the $x-z$ plane for a total state space of each spatio-temporal field of $n= 65792$. Note that the horizontal axis represents the  azimuthal direction ($z$) and the vertical axis represents the radial direction ($x$). The first 20 modes are retained for the training the model.  The associated temporal dynamics are shown in Fig.~\ref{fig:svd2}.}
   \label{fig:svd1}
\end{figure}

\begin{figure*}[t]
   \hspace*{-.7in}
   \begin{overpic}[width=1.2\textwidth]{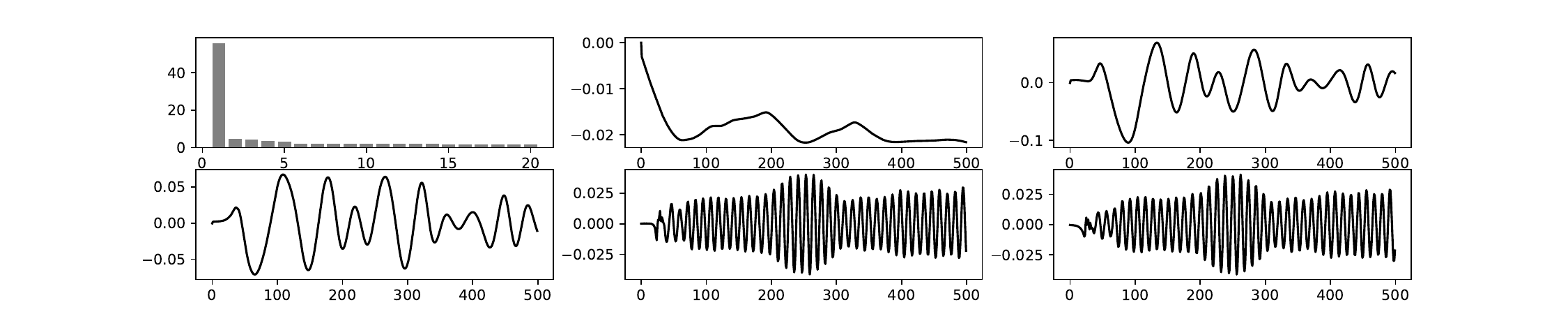}
\put(15,16){(a)}
\put(42.5,16){(b)}
\put(71,16){(c)}
\put(15,7.5){(d)}
\put(42.5,7.5){(e)}
\put(71,7.5){(f)}
\put(7,14){$\frac{100\sigma_j}{\sum \sigma_k}$}
\put(25.4,9.5){$j$}
\put(38,15.4){${\bf v}_1$}
\put(65.2,15.4){${\bf v}_2$}
\put(38,6.4){${\bf v}_4$}
\put(65.2,6.4){${\bf v}_5$}
\put(10,6.4){${\bf v}_3$}
\put(22,-1){time $t$}
   \end{overpic}
   \caption{(a) Singular value ($\sigma$) decay and percentage of variance (${100\sigma_j}/{\sum \sigma_k}$) captured by each mode of the randomized SVD decomposition.  The first 20 modes are retained for SHRED training as these modes capture the dominant activity in the plasma.  The time dynamics of the first 5 modes (${\bf v}_k$ for $k=1, 2, 3, 4, 5$) are illustrated in panels (b)-(f) as a function of time.  The associated spatial modes are shown in Fig.~\ref{fig:svd1}. }
   \label{fig:svd2}
\end{figure*}
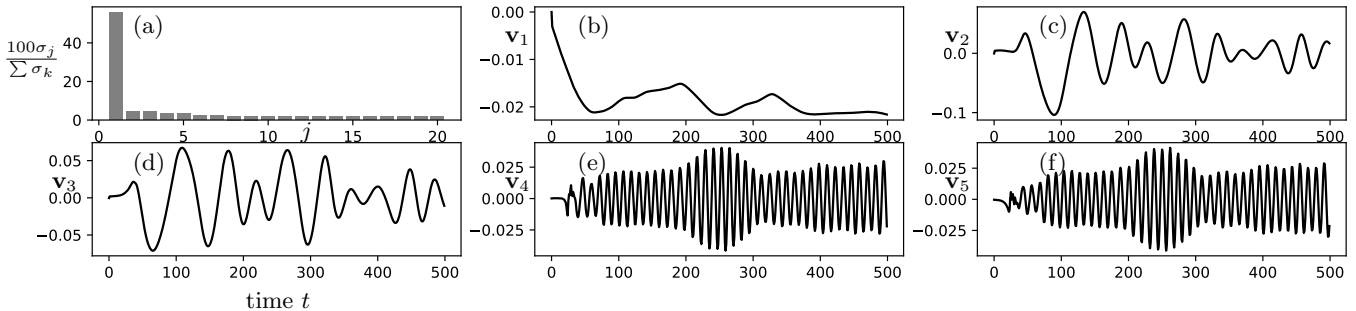

\section{SHRED for Plasma Dynamics}

This section outlines the various steps in building a SHRED model for emulating the dynamics of the ${\bf E} \times {\bf B}$ plasma discharge in the Hall-thruster-representative radial-azimuthal configuration~\cite{reza2023parametric,faraji2023dynamic,faraji2023dynamic2}.   The SHRED architecture leverages compressed representations of the data for efficiency and allows for both reconstruction and forecasting.

\subsection{Compressive Plasma Representation}

The training data are snapshots of the spatio-temporal fields associated with the plasma dynamics.  This includes 14 fields:  ${\bf E}_x$, ${\bf E}_z$, ${\bf J}_{ex}$, ${\bf J}_{ey}$, ${\bf J}_{ez}$, ${\bf n}_e$, ${\bf n}_i$ $\boldsymbol{\phi}$, ${\bf T}_{ex}$, ${\bf T}_{ez}$, ${\bf T}_i$, ${\bf V}_{di,x}$, ${\bf V}_{di,y}$, and ${\bf V}_{di,z}$, which represent, respectively, the radial and the azimuthal electric field components, the radial, axial, and azimuthal electron current densities, the electron and the ion number densities, the azimuthal and the radial electron temperatures, the ion temperature, and the radial, axial and azimuthal ion velocity components. The data has a spatial discretization of $n_x=256$ and $n_z=257$ for a state space of dimension $n= 65792$.  Each spatio-temporal field is saved in a flattened data matrix ${\bf X}^{(k)}\in \mathbb{R}^{n\times m}$ where $m$ are the number of time snapshots and for $k=1,2, \dots 14$.  The full state-space is the collection of all 14 fields which is 921,088 dimensional.  It is noteworthy that the setup of the adopted test case represents only a portion of the entire circumference of a real-size Hall thruster, and further captures a 2D section of an inherently 3D geometry. Thus, for a larger 2D domain or for a real-world 3D geometry, the state-space would have been remarkably higher dimension.
Training on such large data can easily become computationally intractable. However,  we can exploit the low-rank representation of the data to substantially reduce the computational cost of training.

Each data matrix can be decomposed using the singular value decomposition~\cite{trefethen1997numerical,kutz:2013}
\begin{equation}
    {\bf X}^{(k)} = {\bf U}^{(k)} {\bf \Sigma}^{(k)} {{\bf V}^{(k)} }^*
\end{equation}
However, such a computation for very large matrices is expensive computationally.  Instead, one can use randomized algorithms to accurately approximate full SVD decomposition above.  Randomized linear algebra~\cite{halko2011algorithm,drineas2007randomized,erichsonrandomized} has been shown to be an effective method for performing matrix decompositions at scale.  It is based upon sounds mathematical foundations provided the data has a low-rank structure.  The first step in the algorithm is to randomly sample the data with a random test matrix ($\Omega$)
\begin{equation}
    {\bf Y} = {\bf X}^{(k)} {\bf \Omega}
\end{equation}
where ${\bf \Omega}\in \mathbb{R}^{m\times k}$ and
${\bf Y}\in \mathbb{R}^{n\times k}$.  Typically $k\ll m$, which then allows for a rapid ${\bf Q}{\bf R}$ decomposition~\cite{trefethen1997numerical}.  The data matrix ${\bf X}^{(k)}$ is then projected onto the orthogonal columns of the matrix ${\bf Q}$ so that
\begin{equation}
    {\bf B} = {\bf Q}^T {\bf X}^{(k)} \in \mathbb{R}^{k\times m} .
\end{equation}
The SVD is now performed on the much smaller ${\bf B}$ matrix ${\bf B} = \tilde{\bf U}^{(k)} \tilde{\bf \Sigma}^{(k)} { {\tilde{\bf V}}^{(k)*} }$.  Once $\tilde{\bf U}^{(k)}$ is computed, the original high-dimensional SVD componets can be recovered from 
\begin{equation}
    {\bf U}^{(k)} = {\bf Q} \tilde{\bf U}^{(k)} .
\end{equation}
Implementation of randomized linear algebra packages~\cite{halko2011algorithm,drineas2007randomized,erichsonrandomized} are now available in python, matlab or R.  This accelerates the construction of the low-rank matrices 
${\bf U}^{(k)}$, $ {\bf \Sigma}^{(k)}$ and $ {{\bf V}^{(k)} }$ which are used in the archiecture shown in Fig.~\ref{fig:Figure1}.  The computation on a laptop for the given data is approximately 20 minutes without the randomized algorithm and approximately 30 seconds with it.  Thus each data matrix is first processed via the randomized SVD.

\begin{figure*}[t]
   \begin{overpic}[width=0.9\textwidth]{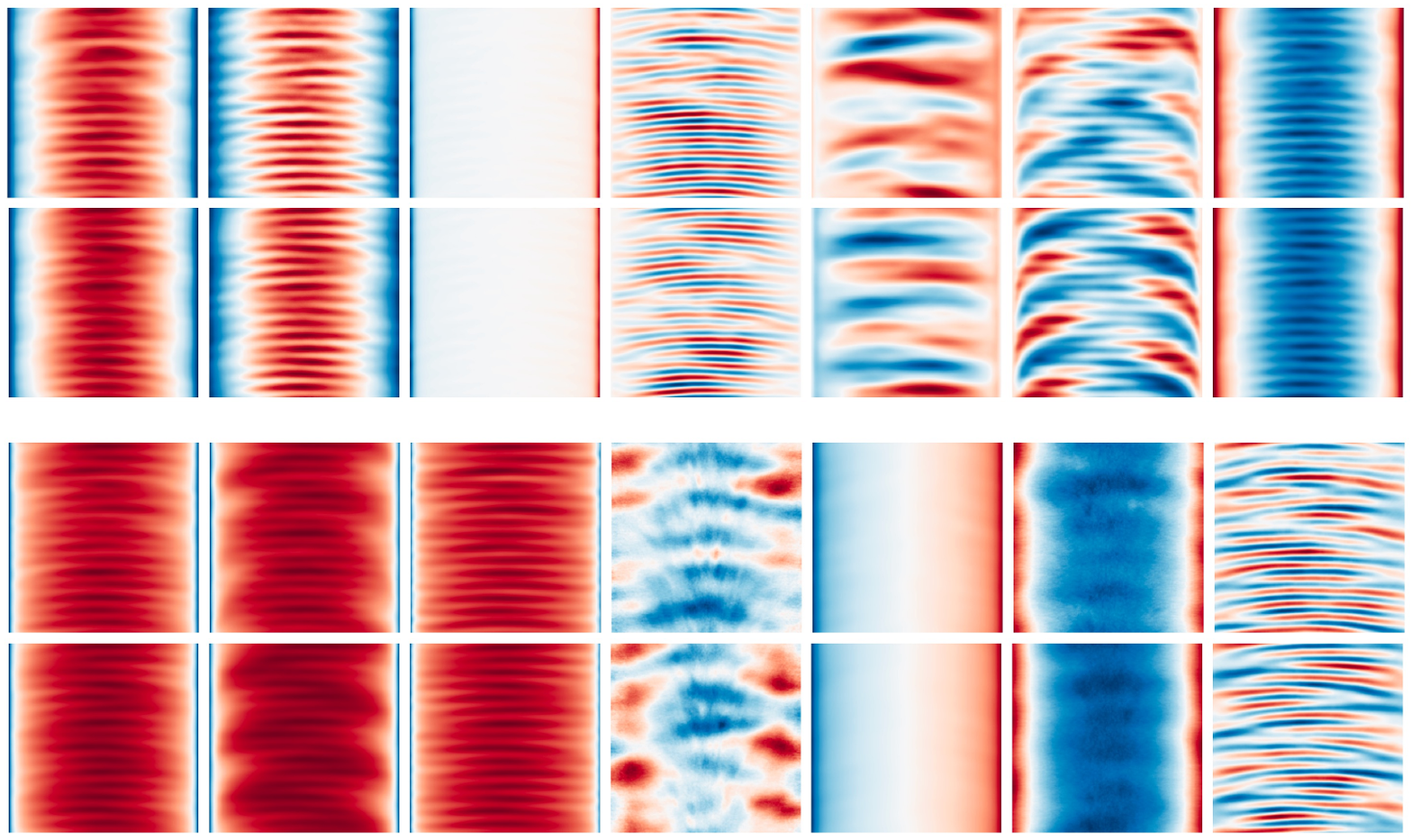}
   \put(2,54){\rotatebox{90}{Truth}}
   \put(2,37){\rotatebox{90}{Reconstruction}}
   \put(2,25){\rotatebox{90}{Truth}}
   \put(2,8){\rotatebox{90}{Reconstruction}}
   \put(10,63){${\bf n}_e$}
   \put(23,63){${\bf n}_i$}
   \put(36,63){${\bf E}_x$}   
   \put(49,63){${\bf E}_z$}   
   \put(62,63){${\bf J}_{ex}$}   
   \put(75,63){${\bf J}_{ey}$}   
   \put(88,63){${\bf J}_{ez}$} 
   \put(10,34.5){$\boldsymbol{\phi}$}
   \put(23,34.5){${\bf T}_{ex}$}
   \put(36,34.5){${\bf T}_{ez}$}   
   \put(49,34.5){${\bf T}_i$}   
   \put(62,34.5){${\bf V}_{di,x}$}   
   \put(75,34.5){${\bf V}_{di,y}$}   
   \put(88,34.5){${\bf V}_{di,z}$}    
   \end{overpic}
   \vspace*{-.4in}
\caption{Reconstruction versus truth for the fourteen spatio-temporal fields of the plasma dynamics.  In this example, the ${\bf n}_e$ field is measured in three randomly locations.  The sensor trajectory is used in SHRED to perform a reconstruction of all fourteen fields.  The truth and reconstructions are shown for withheld test data at a time randomly selected.  As is shown, three point sensor measurements are capable of accurate reconstructions. Note that the horizontal axis represents the azimuthal direction ($z$) and the vertical axis represents the radial direction ($x$).}
   \label{fig:comp1}
\end{figure*}

\begin{figure}[t]
   \begin{overpic}[width=0.45\textwidth]{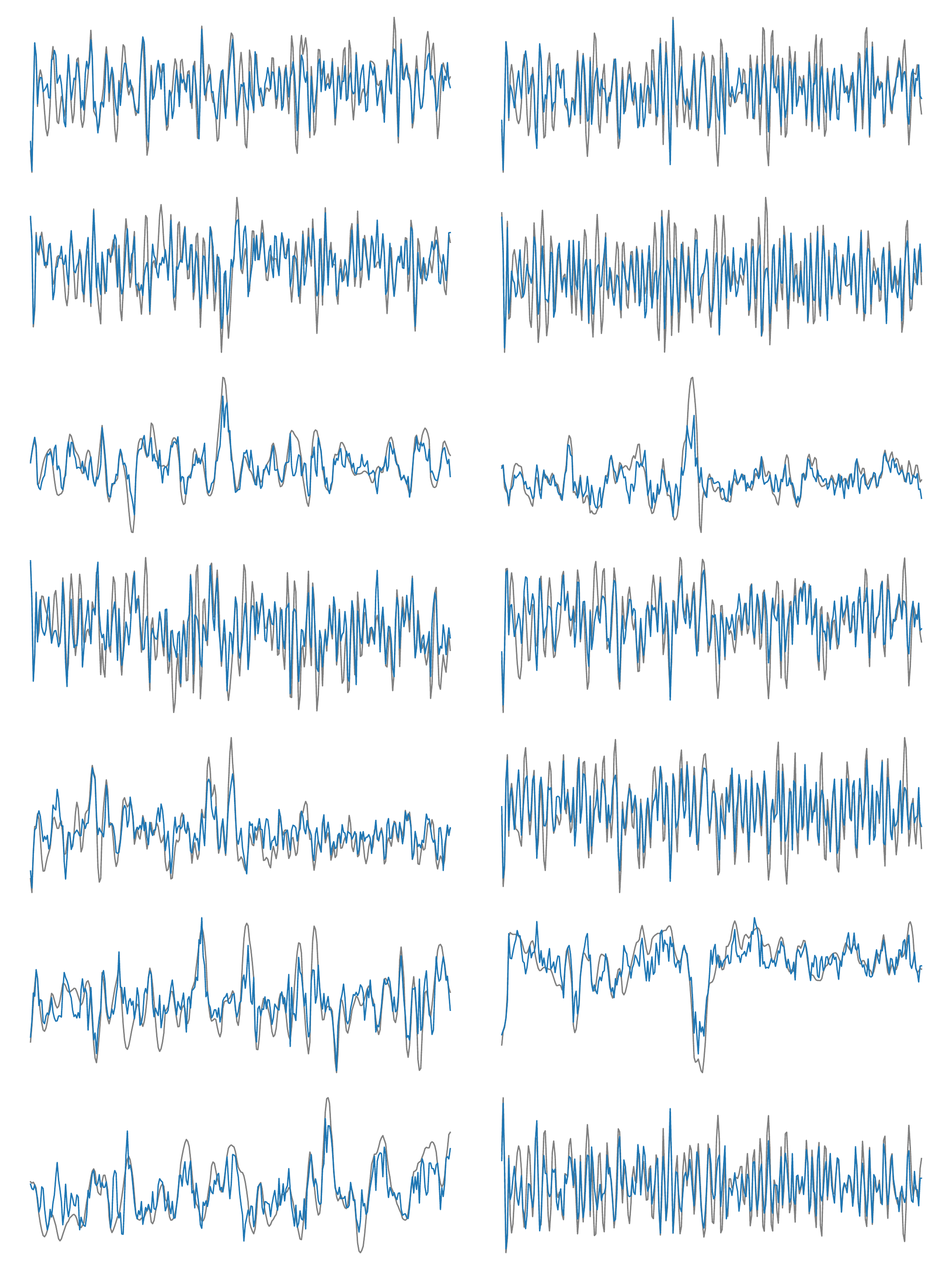}
   \put(2,98){${\bf n}_e$}
   \put(40,98){${\bf n}_i$}
   \put(2,83){${\bf E}_x$}   
   \put(40,83){${\bf E}_z$}   
   \put(2,70){${\bf J}_{ex}$}   
   \put(40,70){${\bf J}_{ey}$}   
   \put(2,55){${\bf J}_{ez}$} 
   \put(40,55){$\boldsymbol{\phi}$}
   \put(2,42){${\bf T}_{ex}$}
   \put(40,42){${\bf T}_{ez}$}   
   \put(2,27){${\bf T}_i$}   
   \put(40,27){${\bf V}_{di,x}$}   
   \put(2,14){${\bf V}_{di,y}$}   
   \put(40,14){${\bf V}_{di,z}$}    
   \put(15,0){time $t$}
   \end{overpic}
\caption{Time dynamics of the fourteen spatio-temporal fields of the plasma dynamics at a randomly selected spatial coordinate.  The blue line is the ground truth and the orange line is the prediction of the temporal dynamics on test data.}
   \label{fig:comp2}
\end{figure}

Figures~\ref{fig:svd1} and \ref{fig:svd2} show the important aspects of the decomposition. Specifically, the first figure shows the first six modes of the SVD computed with the randomized algorithm.  This is for the field variable ${\bf n}_e$.  In Fig. \ref{fig:svd2}, the top left picture shows the singular value decay of the matrix with the first mode dominating the variance.  A total of 20 modes are kept for use in reconstruction and training as shown in Fig.~\ref{fig:Figure1}.  The time dynamics of the first five modes, i.e. the first five columns of $ {{\bf V}^{(k)} }$, are shown in the subsequent panels.  The columns of $ {{\bf V}^{(k)} }$ represents the compressed space on which the SHRED model is trained.

\subsection{Reconstructions}

\begin{figure}[t]
   \begin{overpic}[width=0.45\textwidth]{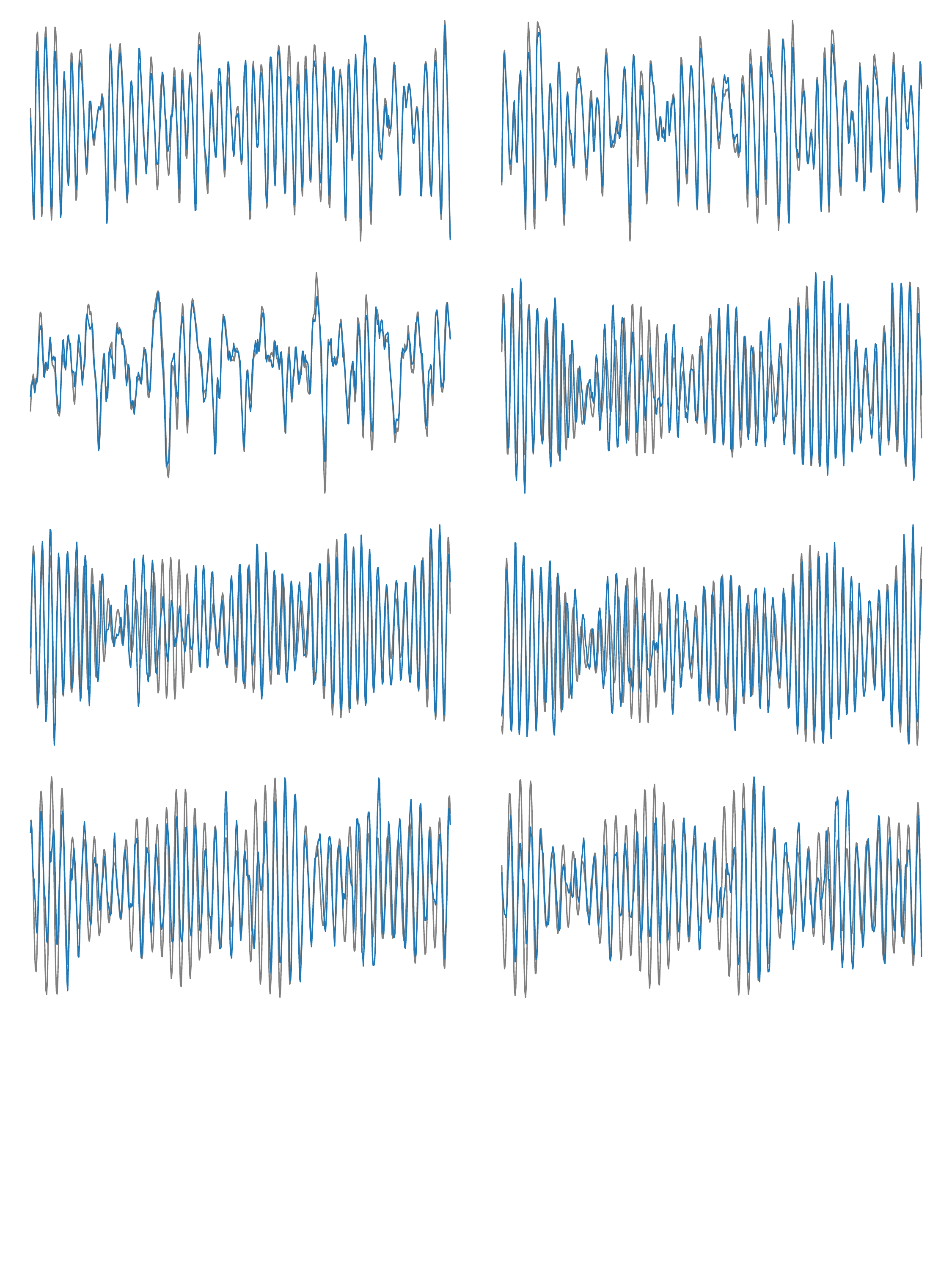}
    \put(2,98.5){${\bf v}_1$}
    \put(2,78){${\bf v}_3$}
    \put(2,60){${\bf v}_5$}
    \put(2,39){${\bf v}_7$}
    \put(40,98.5){${\bf v}_2$}
    \put(40,78){${\bf v}_4$}
    \put(40,60){${\bf v}_6$}
    \put(40,39){${\bf v}_8$}
       \put(15,18){time $t$}
   \end{overpic}
   \vspace*{-.8in}
\caption{Reconstruction (blue) versus truth (gray) for the ${\bf E}_{fZ}$ temporal dynamics of the first eight compressive dynamic components of ${\bf V}^{(k)}$.  The temporal dynamics are used to reconstruct the full spatio-temporal field by projecting back through ${\bf U}^{(k)}$.  }
   \label{fig:fore1}
\end{figure}

\begin{figure*}[t]
   \begin{overpic}[width=0.8\textwidth]{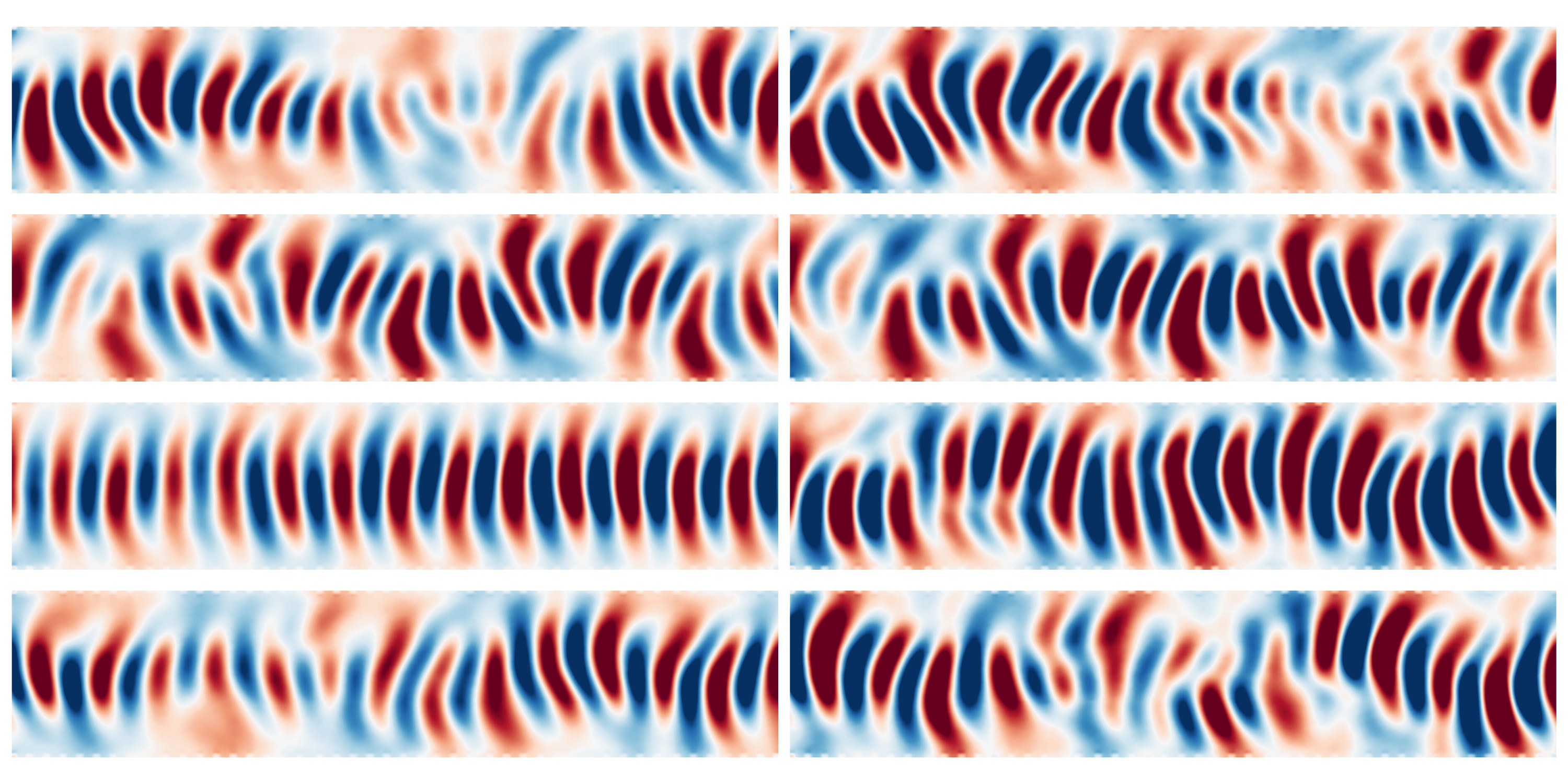}
   \put(-12,42){\rotatebox{0}{${\bf E}_z(400\Delta t)$}}
   \put(-12,30){\rotatebox{0}{${\bf E}_z(800\Delta t)$}}
   \put(-13,19){\rotatebox{0}{${\bf E}_z(1000\Delta t)$}}
   \put(-13,7){\rotatebox{0}{${\bf E}_z(1200\Delta t)$}}
   \put(22,50){Truth}
   \put(70,50){Forecast}
   \end{overpic}
\caption{Reconstruction versus truth for the ${\bf E}_z$ spatio-temporal field of the plasma dynamics forecast from the compressed representation of Fig.~\ref{fig:fore1}.  The comparison is made at $t=400\Delta t, 800\Delta t, 1000\Delta t, 1200\Delta t$ in the future.  The LSTM roll out shows that the SHRED model preserves the fidelity of the high-dimensional simulations. Note that the horizontal axis represents the azimuthal direction ($z$) and the vertical axis represents the radial direction ($x$).}
   \label{fig:fore2}
\end{figure*}

SHRED can be understood as the amalgamation of an LSTM for processing a time-series of sensor measurements followed by a feedforward neural network, or decoder, for reconstructing a high-dimensional state from the learned latent representation of the LSTM \cite{williams2022data}. Let the high-dimensional state to be reconstructed be denoted as $x_T \in \mathbb R ^n$ and assume access to a set of sensor measurements $y_t = C x_t \in \mathbb R ^m$ for $t \in \{T - K, T - K +1, \dots, T -1, T \}.$  $K$ can be determined by empirical analyses of the system at hand.  We assume the measurements are sparse point measurements, that is $ m << n$ and $C$ consists of rows of the $n \times n$ identity matrix, although there is some evidence to suggest that neural network based reconstructions can be performed with nonlinear measurements \cite{erichson2020shallow}.  The set of sensor measurements serve as inputs to an LSTM \cite{hochreiter1997long} with recursive update equations
\begin{gather}
    h _t = \sigma \left( W _o \begin{bmatrix}  h_{t-1}, y_t \end{bmatrix} + b _o \right) \odot \tanh(c _t)\\
    c_t = \sigma \left( W _f \begin{bmatrix}
        h _{t-1}, y_t  
    \end{bmatrix} +b _f \right) \odot c _{t-1} \\ 
    + \sigma \left( W _i \begin{bmatrix}
        h _{t-1}, y_t
    \end{bmatrix} +b _f \right) \odot \tanh \left( W_g \begin{bmatrix}
        h_{t-1}, y_{t}
    \end{bmatrix} + b _g \right) \notag
\end{gather}
where $W _{RN} = \{W _o, W _f, W _i, W _g, b _o, b _f, b _i, b _g \}$ are the trainable weights and biases of the LSTM. We denote
\begin{equation}
    h_T = \mathcal G ( \{ y_t \} _{T - K} ^{T}; W_{RN}).
\end{equation}
The latent state $h_T$ learned by the LSTM has a variety of interesting properties that will be discussed in a later section.  

The feedforward component of the SHRED architecture is a shallow decoder with $b$ layers denoted by
\begin{equation}
    \mathcal F (h; W_{SD} ):= R(W^bR(W^{b-1} \cdots R (W^1h))),
\end{equation}
parameterized by trainable weights $W_{SD} = \{ W^1, \dots, W^b \}$ and with nonlinear scalar activation function $R$ (chosen to be ReLU). In total, the SHRED network is given by 
\begin{equation}
    \mathcal H ( \{ y_t \} _{T - K} ^{T}) =  \mathcal F (\mathcal G ( \{ y_t \} _{T - K} ^{T}; W_{RN}); W_{SD}).
\end{equation}
The network is trained to minimize reconstruction loss over a set of training states $\{x_t \}_{1}^{N}$,
\begin{equation}
    \mathcal{H} \in \mbox{argmin}_{\widetilde{\mathcal{H}} \in  {H}}  \sum _{t=1}^N ||x_i - \widetilde{\mathcal{H}}\left( \{ y _i \}_{i=t-K}^t \right)||_2,
\end{equation}
using the ADAM optimizer \cite{kingma_adam_2017}.  The assumption of access to a set of high-dimensional states for training in this manner is a strong one; simultaneous measurement of an entire high-dimensional system is sometimes simply impossible.  In such cases, a high-fidelity simulation can be used to train the network, provided the simulation accurately approximates the statistics of the real system.  Alternatively, if full-state measurements are possible, but prohibitively expensive in the long-term, the generation of training data can be viewed as a one-time upfront cost.  

Previous work has demonstrated that such networks outperform traditional, POD based techniques for state estimation while requiring fewer available sensors \cite{williams2022data}.  In this work, we consider a time-dependent measurement matrix $C$ that measures only one component of the 14 coupled spatio-temporal fields.
This is in contrast to existing works which can reconstruct only the sensed fields.  We emphasize that while the set of measurement matrices $C$ can be chosen arbitrarily, corresponding training data is necessary in order to train the network.  

SHRED is applied to the radial-azimuthal test-case simulations in order to build a ROM for the plasma dynamics.  The data includes the spatio-temporal dynamics of all fields measured  over a given time course.  For this first example, we demonstrate the reconstruction capabilities of the SHRED architecture. Figure~\ref{fig:comp1} shows a direct comparison for the true dynamics from the PIC simulation versus the SHRED reconstruction on withheld test data from three point sensor measurements of the ${\bf n}_e$ field.  As can be seen in Fig. \ref{fig:comp1}, the SHRED model faithfully reconstructs all fourteen fields after training, even though it only has access to three sensor measurements in the first spatio-temporal field.

The reconstruction of the time dynamics is shown in Fig.~\ref{fig:comp2} for a randomly selected spatial location.  The ground truth and SHRED reconstruction are show
on withheld data.  As can be seen, many of the key and dominant features are retained in the SHRED model in terms of frequency content and intermittent behavior.  The SHRED model however, misses some of the peak values when fitting the ground truth time series data.  Regardless, the quality of reconstruction is quite strong across all fourteen fields, with some better reconstructed than others.

\subsection{Forecasting}

The previous section showed the SHRED model as a reconstruction method, where the training and the test dataset were temporally interspersed. SHRED can instead be made into a forecasting tool by separating the training and the test dataset in a temporally distinct manner. In this way, the train-test split of the data is structured so that the data up to a given time is training data, and all data past that time (future data) is test data.  This sets up SHRED as a roll out model whereby the future states can be predicted by using the recurrent neural network, in this case a LSTM, as the time-sequence prediction engine.  

Predictions are made in the compressive space learned from the ${\bf V}^{(k)}$ found by the singular value decomposition.  Figure~\ref{fig:Figure1} shows that the training is on the right singular values (${\bf V}^{(k)}$) of the matrix decomposition.  The allows for efficient and rapid training.  The left singular vectors (${\bf U}^{(k)}$) are then used to project the data back to the high-dimensional space.  Figure~\ref{fig:fore1} shows the forecast of the first eight modes of ${\bf V}^{(k)}$.  In this example, the SHRED model is trained on the 
${\bf E}_{Z}$ field.  The forecast shows that the SHRED model is especially good at forecasting the most dominant modes, which is important for reconstruction as these contribute the most to the overall spatio-temporal data.  The reconstructions in the high-dimensional space are compared to the ground truth in Fig.~\ref{fig:fore2} for four different future time points $t=400\Delta t, 800\Delta t, 1000\Delta t, 1200\Delta t$.  Again, the performance is quite strong and model is able to accurately predict the spatio-temporal dynamics of the ${\bf E}_{Z}$ field.  Importantly, the LSTM roll out provides a model that preserves the structure of the underlying dynamical system encoded in the compressed temporal representation.

\section{Conclusion}

We have demonstrated a novel architecture for reduced order modeling in plasma physics.  This model is the first to leverage the temporal encoding in the sensor measurements for recreation of spatial behavior.  Rooted in the theory of separation of variables, the method generalizes to a nonlinear setting where the time encoding is done with a recurrent neural network and the spatial decoding is performed with a shallow decoder network. SHRED offers a lightweight model that only requires access to three sensor measurements of a single field in order to reconstruct the global dynamics for all 14 fields describing the evolution of an ${\bf E}\times {\bf B}$ Hall-thruster-representative plasma.

The origin of the three required sensor measurements is conjectured to be rooted in the triangulation of sources and dynamics.  Similarly to three cell phone towers which are required to localize a source, so do the sensors uniquely identify and learn the patterns of dynamics across the domain.  Indeed, three sensors performs as well as 10, 20 or more sensors since it has all the information required in the time sequence for producing accurate full state reconstructions.   The sensors are chosen randomly.  And with high-probability, any three random configurations are as good as any other.  Of course, edge cases where sensors are essentially all picked in the same location or vicinity can be problematic, but this is an extremely low-probability event and can be easily checked.  Alternatively, there may be regions in the domain that are {\em deep spots} dynamically so that the sensor gains no information.  Such locations also should be disallowed.    

As a reduced order modeling paradigm, the SHRED model is a compact and lightweight encoding of the full spatio-temporal dynamics.  It thus allows for rapid emulation and roll outs at significant computational cost savings relative to numerical simulations.  As such, this provides a viable emulation tool for encoding the physics of the plasma. More than that, it provides a potential architecture for improving models by integrating the trained SHRED model directly within experiment.  Specifically, one can imagine direct measurement of the ${\bf n}_e$ field, e.g., at a discrete number of locations within the Hall thruster's domain.  The trained model can then produce a prediction of the remaining field components.  This can be used to evaluate whether the predicted fields are viable physically.  In addition, the predictions can be used to suggest updates to the simulation model for the plasma itself and to quantify the discrepancy between the simulation and the experiment, hence, identifying the areas that need improvement in the simulation for more accurate numerical results. Thus, the interplay of experiment and simulation can be leveraged fully with the SHRED model towards building better theoretical understanding and models of plasmas.  This will be pursued in future work.

Ultimately, envisioning the realization of digital twins as transformative tools to empower science and engineering behind Hall thrusters and the broader family of plasma technologies and systems, the SHRED architecture presents itself as an optimal joint platform of simulation models and measurements data, providing a seamless integration, data interfaces and pipeline. This suggests that SHRED has a great potential to serve as an enabler of the digital twin technology for plasmas.

\section*{Code:}  The code and data (compressed) is avaialble at:
\url{https://github.com/nathankutz/plasmaSHRED}

\section*{Acknowledgments} The work of JNK was supported in part by the US National Science Foundation (NSF) AI Institute for Dynamical Systems (dynamicsai.org), grant 2112085.

\bibliographystyle{unsrt}
\bibliography{bibliography,references,merged}

\end{document}

%% file: z_vardef.tex
\DeclareMathAlphabet{\mathpzc}{OT1}{pzc}{m}{it}
 